\documentclass[sigconf]{acmart}
\usepackage{algorithm}
\usepackage{algorithmic}
\usepackage{multirow}
\usepackage{colortbl} 
\usepackage{subfigure}
\AtBeginDocument{%
  }

\copyrightyear{2026}
\acmYear{2026}
\setcopyright{cc}
\setcctype{by-nc-nd}
\acmConference[MM '26]{Proceedings of the 34th ACM International Conference on Multimedia}{November 10--14, 2026}{Rio de Janeiro, Brazil}
\acmBooktitle{Proceedings of the 34th ACM International Conference on Multimedia (MM '26), November 10--14, 2026, Rio de Janeiro, Brazil}
\acmDOI{10.1145/3767308.3835793}
\acmISBN{979-8-4007-2213-4/2026/11}

\begin{document}

\title{D3ER: Supporting Multi-Modal Recommendation via Disentangle and Distillation-based Dynamic Ensemble}


\author{Bingnan Wang}
\authornote{Both authors contributed equally to this work.}
\affiliation{
  \institution{Institute of Software, Chinese Academy of Sciences}
  \city{Beijing}
  \country{China}
}
\affiliation{
  \institution{University of Chinese Academy of Sciences}
  \city{Beijing}
  \country{China}
}
\email{wangbingnan21@mails.ucas.ac.cn}

\author{Yi Li}
\authornotemark[1]
\affiliation{
  \institution{Institute of Software, Chinese Academy of Sciences}
  \city{Beijing}
  \country{China}
}
\affiliation{
  \institution{University of Chinese Academy of Sciences}
  \city{Beijing}
  \country{China}
}
\email{liyitunan@gmail.com}

\author{Xiongxin Tang}
\affiliation{
  \institution{Institute of Software, Chinese Academy of Sciences}
  \city{Beijing}
  \country{China}
}
\affiliation{
  \institution{University of Chinese Academy of Sciences}
  \city{Beijing}
  \country{China}
}
\email{xiongxin@iscas.ac.cn}

\author{Fanjiang Xu}
\affiliation{
  \institution{Institute of Software, Chinese Academy of Sciences}
  \city{Beijing}
  \country{China}
}
\affiliation{
  \institution{University of Chinese Academy of Sciences}
  \city{Beijing}
  \country{China}
}
\email{fanjiang@iscas.ac.cn}

\author{Jiangmeng Li}
\correspondingauthor
\affiliation{
  \institution{Institute of Software, Chinese Academy of Sciences}
  \city{Beijing}
  \country{China}
}
\affiliation{
  \institution{University of Chinese Academy of Sciences}
  \city{Beijing}
  \country{China}
}
\email{jiangmeng2019@iscas.ac.cn}

\renewcommand{\shortauthors}{Bingnan Wang, Yi Li, Xiongxin Tang, Fanjiang Xu, and Jiangmeng Li}

\begin{abstract}
Incorporating items' information shared among multiple modalities into a fused representation, multi-modal recommendation (MR) has demonstrated documented success than canonical unimodal recommendation. Although several attempts have been made to extract the discriminative information unique in each modality, existing methods suffer from a core limitation: the joint learning of \textit{modal-homogeneity discriminative information} (HOI) and \textit{modal-heterogeneity discriminative information} (HEI) tends to weaken their individual effectiveness. To remedy this deficiency, we propose a novel method, dubbed Disentangle and Distillation-based Dynamic Ensemble for multi-modal Recommendation (D3ER). We introduce gradient boosting into MR for the first time to formalize the optimization objective for alternately learning HOI and HEI. This design enables models dedicated to each type of information to focus on their proficient samples, thereby promoting specialized optimization. Furthermore, to mitigate the inherent high storage cost and risk of local optima in gradient boosting, we enhance our framework with knowledge distillation and a global correction regularization. Experiments on prevalent real-world datasets confirm the superiority of our proposed method on MR.
\end{abstract}



\maketitle

\section{Introduction}
\label{sec:intro}
Recommendation systems have attracted increasing research attention due to their broad application in e-commerce, social media, etc.
It aims to recommend items of interest to users based on historical interactions. With the popularity of multimedia platforms, multi-modal recommendation \cite{MMGCN, BM3} incorporates rich semantic content of items (e.g., visual appearance and textual descriptions) to compensate for the foundational collaborative filtering (CF) \cite{he2017neural} framework, thus capturing preferences of users more accurately. 

\begin{figure*}[htbp]
\centering
\includegraphics[width=0.85\linewidth]{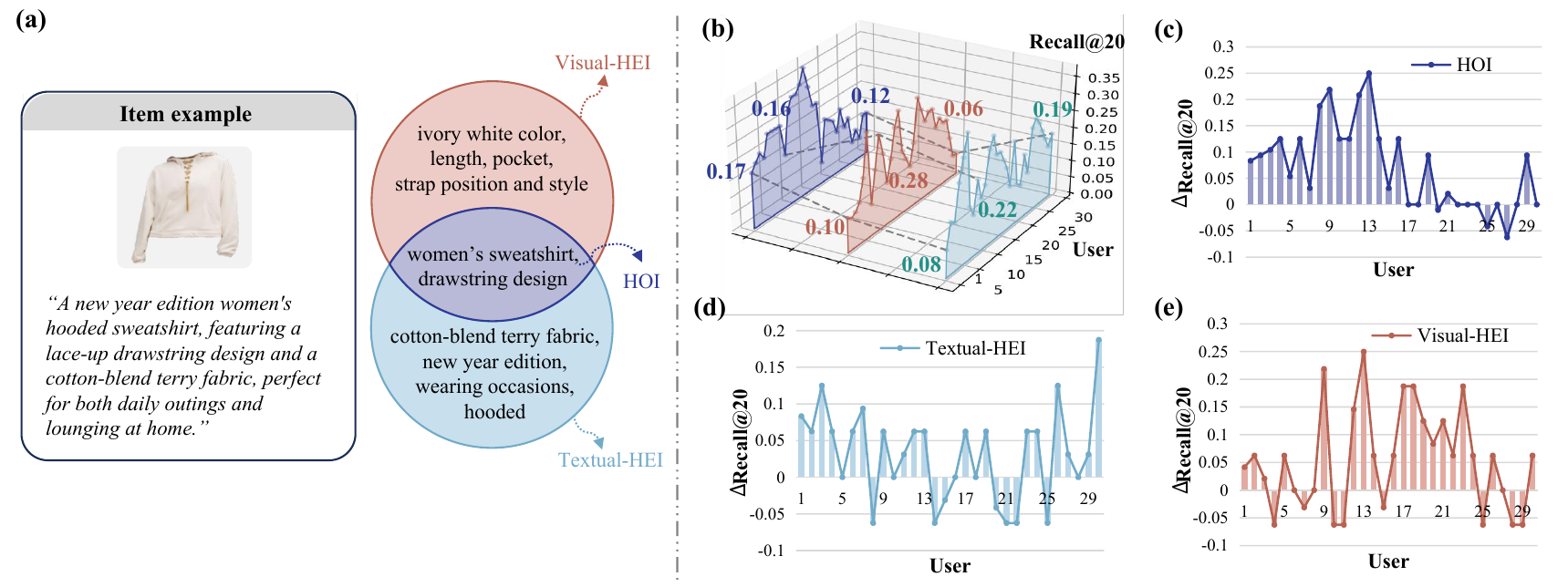}
\vspace{-0.2cm}
\caption{(a) illustrates the HOI and HEI of an item example. (b) shows the prediction preference of HOI and HEI for thirty randomly selected users of Baby dataset \cite{Amazon}. HOI and HEI are obtained through the proposed module detailed in \textbf{Section} \ref{sec:FCD}. (c-e) The performance difference of independent and joint optimization for three types of information.}
\label{fig_mtv}
\vspace{-0.2cm}
\end{figure*}

The existing paradigm for MR typically adheres to the following pipelines \cite{wei2020graph, wang2021dualgnn, liu2023semantic, zhou2023tale, guo2024space}:
(\textit{\textbf{i}}) Modality-specific feature extractors are employed to derive the initial representations for each modality; (\textit{\textbf{ii}}) Modality-specific graph convolutional networks (GCNs) \cite{LightGCN} utilize the initial modality representation as node embeddings to inject high-order collaborative signals into the refined representation;
(\textit{\textbf{iii}}) The refined embeddings of each modality are fused together \cite{MMGCN} to generate the final user and item representations. To alleviate the inherent data sparsity in historical interactions, these methods typically assume that \textit{modal-homogeneity discriminative information} (HOI), shared across modalities, is beneficial to recommendation performance, thereby imposing additional modality alignment self-supervised signals. However, such constraints exclude the unique semantics embodied in \textit{modal-heterogeneity discriminative information} (HEI), hindering the potential discriminability of MR systems.

Figure \ref{fig_mtv}(a) showcases an example of HOI and HEI in MR. 
Unique information, such as the fabric material in the textual modality, will also affect the decision of users. To address this challenge, some studies explore modality-specific information \cite{han2022modality,he2024never}. But these works typically adopt a naive ensemble strategy that jointly learns homogeneous and heterogeneous information.
Such a strategy tends to dilute the sample-oriented predictive preferences, thereby undermining the unique advantages of these information. Figure \ref{fig_mtv}(b) demonstrates the prediction preference of HOI or HEI across diverse samples. It can be seen that HOI and HEI exhibit distinct contributions that vary with respect to users.

On top of this, we observe that the joint optimization strategy diminishes the discriminative power in learned representations based on the results in Figure \ref{fig_mtv}(c-e). Specifically, we report the performance differences $\Delta R =R_{id}-R_{jt}$, where $R_{id}$ denotes the Recall@20 obtained by independent optimization, and $R_{jt}$ represents that of joint learning. The consistently inferior performance of joint optimization indicates that the naive ensemble degrades the discriminative ability of each type of information. This gives rise to a critical question: \textit{how to develop an effective ensemble strategy to obtain comprehensive sample-oriented discriminative information?}

To this end, we propose \textit{\textbf{D}isentangle and \textbf{D}istillation-based \textbf{D}ynamic \textbf{E}nsemble for multi-modal \textbf{R}ecommendation} (\textbf{D3ER}), a novel framework to facilitate the comprehensive learning of HOI and HEI. In D3ER, we extract modality-shared and modality-specific features through the feature component disentanglement (FCD) module under the constraints of inter-modal alignment and intra-modal separation. To alleviate the distribution misalignment problem caused by contrastive learning, our FCD performs alignment at both the instance level and distribution level. Subsequently, we propose a knowledge distillation-enhanced gradient boosting (KDBoost) module to learn comprehensive sample-oriented discriminative information. To make up for the weakness of naive ensemble, we introduce gradient boosting \cite{chen2016xgboost} into MR for the first time. The optimization objective of gradient boosting is formulated to fit the negative gradient of the task-specific loss with respect to the previous cumulative model. This design leads each model to focus on different subsets of samples, aligning well with our motivation to address the weakening of the sample-oriented prediction preference. However, traditional gradient boosting relies on retaining all historical weights to compose a robust overall model \cite{chen2016xgboost}, which is infeasible in resource-constrained scenarios. To this end, we incorporate a distillation step in KDBoost to reduce storage cost while maintaining the discriminative information .Besides, an extra global correction regularization is introduced to alleviate the risks of greedy learning inherent in boosting. Extensive experiments on three real-world datasets demonstrate the superiority of D3ER, which is further supported by theoretical analyses that establish a tighter error bound for the method. 

Our main contributions can be encapsulated as follows: 
\begin{itemize}
\item We illustrate the challenge of weakening the sample-oriented prediction preference in the joint learning of HOI and HEI.

\item We propose D3ER, which integrates the gradient boosting for the first time to facilitate the comprehensive learning of sample-oriented discriminative information for both HOI and HEI.

\item We design a feature component disentanglement module to obtain HOI and HEI by performing the inter-modal alignment and intra-modal separation.

\item Empirical evaluations on prevalent MR datasets substantiate the superior performance of our method. Additionally, we provide theoretical guarantees on our proposed method.
\end{itemize}

\section{Related Work}

\subsection{Multi-modal Recommendation}
How to effectively model user profiles by leveraging interactions and modality features poses challenges to MR. Researchers have designed various auxiliary graphs, such as user or item co-occurrence graphs \cite{wang2021dualgnn,zhou2023tale}, hypergraphs \cite{guo2024lgmrec, zhang2022price}, and generated graphs \cite{MMSSL, DiffMM}, to complement original user-item bipartite graphs. Another line of research focuses on introducing self-supervised signals \cite{ren2024sslrec, SLMRec} to mitigate the sparsity of the BPR loss \cite{MF-BPR}. Typically, these approaches align representations across modalities by imposing contrastive loss \cite{MMGCL, MMSSL}. Although earlier works \cite{han2022modality,wang2021multimodal} attempt to incorporate the idea of disentangled representation learning into MR, they still encounter some defects. PAMD \cite{han2022modality} relies on Euclidean distance to obtain modality-shared information, which could lead to feature collapse \cite{wang2020understanding}, thus resulting in insufficient disentanglement. Besides, the disentangled representations are not guaranteed to be inherently consistent with the recommendation task \cite{wang2022chaos}. MDR \cite{wang2021multimodal} conducts disentanglement in a weak supervision manner, which necessitates the annotation of item attributes. However, the proposed D3ER can conduct the disentanglement in an unsupervised or self-supervised manner and employs instance-level InfoNCE loss and distribution-level loss for more sufficient disentanglement. Moreover, these methods rely on the naive ensemble, resulting in suboptimal exploitation of sample-specific homogeneous and heterogeneous information, whereas our D3ER employs alternate learning to mitigate this issue.

\subsection{Gradient Boosting in Deep Networks}
Gradient boosting is a widely adopted algorithm in ensemble learning. Freund et al. \cite{freund1997decision} propose AdaBoost for classification tasks, where multiple weak learners are trained sequentially to fit the residuals of previous learners. GBDT \cite{friedman2001greedy} extends AdaBoost to a more general context. With the strength of deep learning, numerous efforts have been made to integrate ensemble learning with neural networks. AdaNet \cite{cortes2017adanet} proposes adaptive learning of the network structure using network layers as weak learners. BoostResNet \cite{huang2018learning} explores the ResNet-style architecture by sequentially training multiple residual layers. 
AdaGCN \cite{sun2019adagcn} incorporates AdaBoost into the design of deep graph convolutional networks to extract knowledge from different hops of neighbors. 
Lu et al. \cite{DBLP:conf/icml/LuB0XW24} incorporate ensemble learning into the zero-shot generalization task of multiple pre-trained vision-language models, leveraging knowledge from models of varying scales. 
In our paper, we introduce gradient boosting enhanced by knowledge distillation to facilitate the learning of comprehensive sample-oriented discriminative knowledge for HOI and HEI.

\begin{figure*}[htbp]
\centering
\includegraphics[width=0.8\linewidth]{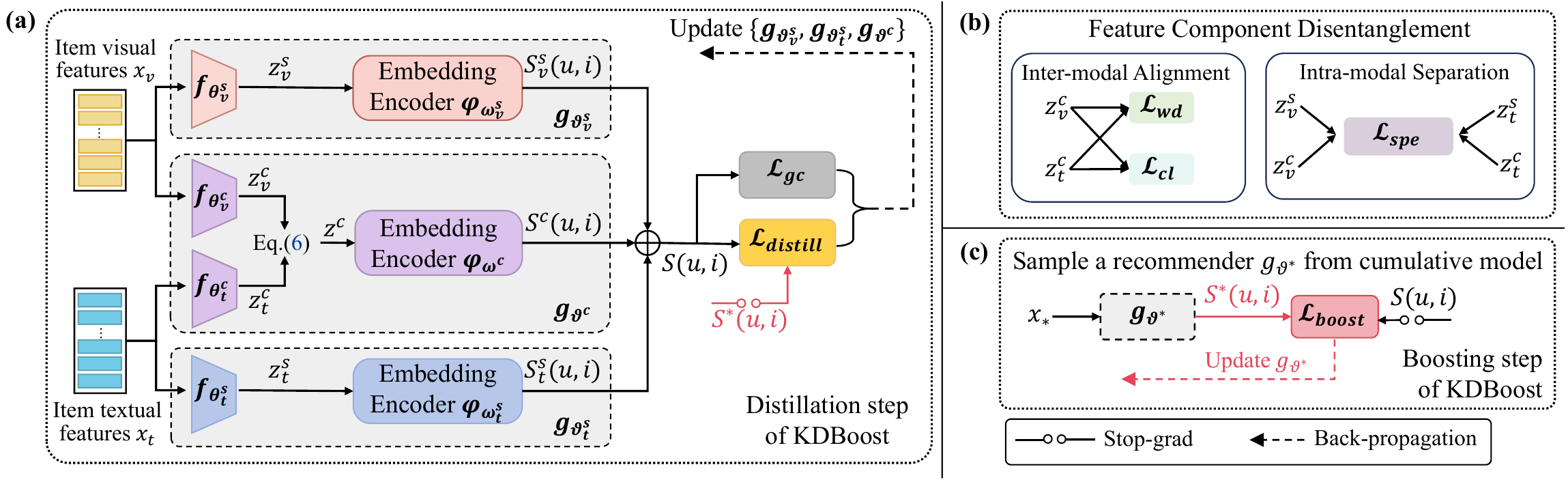}
\vspace{-0.2cm}
\caption{The framework of our D3ER. (a) and (c) show the two steps of KDBoost. (b) gives an illustration of the FCD module.
}
\vspace{-0.2cm}
\label{fig_method}
\end{figure*}

\section{Methodology}
\label{method}
In this section, we first present the preliminaries, including the formulation of the MR task, the definitions of HOI and HEI, and the outline of our proposed D3ER framework. Then we elaborate on each designed module in our method separately.

\subsection{Preliminaries and Overview}

\subsubsection{Task Formulation} 

Given a set of users $\mathcal{U} $, a set of items $\mathcal{I}$, a user-item interaction matrix $\mathcal{A} \in \left \{0,1\right \} ^{\left | \mathcal{U} \right |\times \left | \mathcal{I} \right |}$, and raw modality features $\left \{ x_m \in \mathbb{R}^{d_m} |m\in\mathcal{M}  \right \} $ extracted from pre-trained modality encoders, the MR model aims to learn the representation of users $\mathbf{e}_u \in \mathbb{R}^d $ and items $\mathbf{e}_i \in \mathbb{R}^d $, then obtains similarity scores via inner product, that is, $S(u,i)=\mathbf{e}_u \cdot  \mathbf{e}_i^T$. During inference, the top-$n$ items with the highest scores will be recommended to the user. In this paper, we mainly consider visual and textual modalities (i.e., $\mathcal{M}=\left \{ v,t \right \} $), which is a mainstream configuration in MR.

Given $i \in \mathcal{I}$, the multi-modal contents $x^i_v, x^i_t$ and the corresponding recommendation label $y$, we can formalize the definition of HOI and HEI from the information theory \cite{ash2012information} perspective.

\begin{definition} (\textbf{HOI)} \textit{HOI is the task-dependent information shared by both multi-modal contents:} $HOI = I(x^i_v;x^i_t;y)= I(x^i_v;x^i_t)- I(x^i_v;x^i_t|y)$.
\end{definition}

\begin{definition} (\textbf{HEI}) \textit{HEI is the modal-specific task-dependent information in each unimodal content:} $HEI(x^i_v) = I(x^i_v;y | x^i_t)$ and $HEI(x^i_t) = I(x^i_t;y | x^i_v)$.
\end{definition}

\subsubsection{Overview} The framework diagram of D3ER is illustrated in Figure \ref{fig_method}, which primarily consists of two ingredients: the feature component disentanglement (FCD) module and the knowledge distillation-enhanced gradient boosting (KDBoost) module. The FCD module acquires modal-homogeneity and modal-heterogeneity item representations by inter-modal alignment and intra-modal separation. Then, the KDBoost module is designed to facilitate the comprehensive learning of HOI and HEI through a boosting step and a distillation step.

\subsection{Feature Component Disentanglement}
\label{sec:FCD}
For each modality, we leverage two distinct feature extractors to map the raw item features into two components $z_m^c$ and $z_m^s$, which are denoted by:
\begin{equation}
    z_m^c=f_{\theta_m^c }(x_m), z_m^s=f_{\theta_m^s }(x_m).
\end{equation}
Both $f_{\theta_m^c }(\cdot)$ and $f_{\theta_m^s }(\cdot)$ are implemented by MLPs.

\subsubsection{Inter-Modal Alignment} To ensure that $z_m^c$ encompasses the HOI component of modal $m$, consistent with \cite{he2024never}, we enforce InfoNCE loss \cite{chen2020simple, he2020momentum} to conduct alignment on $\{(z_{m_1}^c,z_{m_2}^c)|m_1, m_2 \in \mathcal{M}\wedge m_1 \neq m_2\} $. Concretely, representations from different modalities of the same item form positive pairs, while those from different items serve as negative pairs:
\begin{equation} \label{eq:contrast}
    \mathcal{L}_{cl}= \sum_{\substack{m_1,m_2\in\mathcal{M}}}^{m_1\ne m_2} \sum_{i\in\mathcal{I}}^{} \frac{exp({z_{m_1,i}^{c}}  {z_{m_2,i}^{c}}^T/\tau )}{\sum_{j\in\mathcal{I}}^{}\sum_{m\in\mathcal{M}}^{}  (exp({z_{m_1,i}^c} {z_{m,j}^c}^T/\tau ))},
\end{equation}
where $\tau$ denotes the temperature parameter. However, although the InfoNCE helps in the coherence between the learned representations and downstream tasks \cite{wang2022chaos}, it suffers from the notorious issue of false negative samples \cite{huynh2022boosting,zhao2021graph}. Consequently, relying solely on instance-level alignment results in $z_m^c$ ($m \in \mathcal{M}$) with identical semantic information fitting diverse distributions in the latent space, limiting the exploration of HOI.

Therefore, in addition to the instance-level loss $\mathcal{L}_{cl}$, we introduce a complementary distribution-level alignment term for $z_m^c$ ($m \in \mathcal{M}$). Specifically, assuming $z_m^c \sim \mathcal{P}(z_m^c)$, we minimize the discrepancy between distributions $\mathcal{P}(z_m^c)$ across modalities using the Wasserstein distance \cite{villani2009wasserstein}, which provides a complete distribution distance metric (i.e., it satisfies non-negativity, symmetry, and the triangle inequality). Let $\mathcal{P}$ be Borel probability measures with finite p-th moment, and for  $\mathcal{P}_1,\mathcal{P}_2 \in \mathcal{P}$, we have the corresponding support sets $\sigma_r,\sigma_g$ and $x_1 \in \sigma_r, x_2 \in \sigma_g$. The $p$-Wasserstein distance of $\mathcal{P}_1,\mathcal{P}_2$ is defined as:
\begin{equation}
W_{p}\left(\mathcal{P}_{1}, \mathcal{P}_{2}\right)=\left(\inf _{\mu \in \Gamma\left(x_{1}, x_{2}\right)} \int d\left(x_{1}, x_{2}\right)^{p} d \mu(x_1, x_2)\right)^{\frac{1}{p}}.
\end{equation}

However, calculating $W_p$ directly is problematic because of the expensive computational costs. 
To achieve a precise estimation with limited time complexity, we calculate Wasserstein distance by a linear integral method based on the Radon transform, which is implemented by: 
\begin{equation}
\begin{aligned}
&\mathcal{W}_{p}(\mathcal{P}_1, \mathcal{P}_2)=\left(\int_{\mathbb{S}^{d_{\theta}-1}} W_{p}\left(\mathcal{R}\mathcal{P}_1(\cdot, \theta), \mathcal{R}\mathcal{P}_2(\cdot, \theta)\right) d \theta\right)^{\frac{1}{p}},
\end{aligned}
\end{equation}
where $\mathcal{R}$ is the standard Radon transform. Concretely, the loss function of distribution-level alignment can be formulated as:
\begin{equation}
    \mathcal{L}_{wd} = \sum_{m_1,m_2 \in \mathcal{M}}^{m_1 \neq m_2}
    \mathcal{W}_{p}\left(\mathcal{P}\left(z_{m_1}^c\right), \mathcal{P}\left(z_{m_2}^c\right)\right).
\end{equation}

Moreover, we utilize the Wasserstein barycenter to obtain HOI of $m$ modalities following \cite{cao2022otkge}:
\begin{equation} \label{eq:wdbf}
    z^c = \mathop{min}\limits_{z} \sum_{m \in \mathcal{M}} \mathcal{W}_p(\mathcal{P}(z_m^c),\mathcal{P}(z)).
\end{equation}

\subsubsection{Intra-Modal Separation} To guarantee that the extracted $z_m^s$ captures HEI, we impose constraints to make them dissimilar to $z_m^c$. Inspired by \cite{dong2023simmmdg}, we introduce the disparity loss to enforce intra-modality separation, utilizing the $\mathcal{L}_2$ distance to measure dissimilarity, which can be written as:
\begin{equation}
    \mathcal{L}_{sep}=-min(\sum_{m\in\mathcal{M} }^{} \sum_{i\in\mathcal{I}}^{}\left \| z_{m,i}^c-z_{m,i}^s \right \|_2^2,d_m ),
\end{equation}
where $d_m$ refers to the distance threshold for modality $m$ and functions as a hyper-parameter to mitigate the risk of excessive learning of task-irrelevant information. Overall, the optimization objective of the feature component disentanglement module is a combination of three terms:
\begin{equation}
\label{eq:fcd}
    \mathcal{L}_{FCD}=\alpha_c\mathcal{L}_{cl}+ \alpha_w\mathcal{L}_{wd}+\mathcal{L}_{sep},
\end{equation}
where $\alpha_c$ and $\alpha_w$ are incorporated for balancing the importance of the instance-level and distribution-level alignment.

\subsection{Knowledge Distillation-enhanced Gradient Boosting} 
\label{sec:kdboost}
We build recommendation models for each type of information (i.e., HOI, visual-HEI, and textual-HEI), thereby forming a cumulative model, i.e., $\mathcal{G}_\Theta =\left \{ g_{\vartheta^c}, g_{\vartheta_v^s}, g_{\vartheta _t^s} \right \} $.
Each model $g_{\vartheta^*}\in\mathcal{G}$ comprises corresponding feature extractors $f_{\theta ^*}(\cdot)$ and an embedding encoder $\varphi_{\omega^*}(\cdot)$, as shown in Figure \ref{fig_method}.
The embedding encoder can be any GCN-based CF model, with the features extracted by the FCD module serving as item node embeddings. The cumulative model obtains the final score by summing up the scores predicted by each model:
\begin{equation}
    S(u,i)= \mathcal{G}_{\Theta } (x,u,i)=  \sum_{g_{\vartheta}\in \mathcal{G}_{\Theta}}^{} g_\vartheta (x,u,i).
\end{equation}

As a straightforward counterpart to our alternating design, a naive ensemble strategy would jointly optimize the three recommenders using the BPR loss, which is expressed as:
\begin{equation}
\label{eq:bpr}
    \begin{aligned}
    &\mathcal{L}_{BPR}=-\sum_{(u,i,j)\in\mathcal{O} }^{} log(\sigma (\bigtriangleup S(u,i,j) )), \\ 
    & \bigtriangleup S(u,i,j)= S(u,i)-S (u,j), 
    \end{aligned}
\end{equation}
where $\mathcal{O}$ is the set of triplets $(u,i,j)$ in dataset, indicating an observed interaction on $(u,i)$ and an unobserved interaction between $(u,j)$, $\bigtriangleup S$ refers to the score difference between pairs $(u,i)$ and $(u,j)$, and $\sigma(\cdot)$ is the sigmoid function. However, as discussed in Section \ref{sec:intro}, the naive ensemble strategy cannot learn comprehensive sample-oriented discriminative information. 

In contrast, our KDBoost module adopts an alternating learning strategy to explore HOI and HEI, thus boosting the ability of each recommender.
The optimization goal of KDBoost is to learn the residual discriminative information of samples that the previous cumulative model has not learned. In this way, each model is encouraged to focus on the samples whose residual discriminative information aligns with its own strengths, rather than uniformly fitting all samples.

\subsubsection{Boosting Step} Inspired by gradient boosting \cite{friedman2001greedy}, we measure the residual discriminative information by the negative gradient of the task loss function with respect to the previous cumulative model. The current model is trained to fit the negative gradient, denoted by:

\begin{equation}
    \begin{aligned}
        \mathcal{L}^{'}_{boost} &= \sum_{(u,i,j)\in\mathcal{O}}^{} \left( \bigtriangleup S_k^* -\left(-\frac{\partial \mathcal{L}_{BPR}(\texttt{sg}(\bigtriangleup S_{k-1} )) }{\partial \bigtriangleup S_{k-1} } \right) \right)^2   \\
        &= \sum_{(u,i,j)\in\mathcal{O}}^{} \left ( \bigtriangleup S_k^* - \left(1-\sigma(\texttt{sg}(\bigtriangleup S_{k-1})) \right)  \right )^2,  
    \end{aligned}
\end{equation}
where $\texttt{sg}()$ means stop gradient, $\bigtriangleup S_k^*$ represents the score difference predicted by the current model $g_{\vartheta^*}$ of epoch $k$, and $\bigtriangleup S_{k-1}$ denotes the score difference predicted by the previous cumulative model of epoch $k-1$,  $(u, i,j)$ is omitted for brevity. Note that for the initial epoch, i.e., $k=1$, the original BPR loss is utilized to train the initial model. Overall, the optimization objective of the boosting step can be formalized as:
\begin{equation}
\label{eq:boost}
    \mathcal{L}_{boost}=\left\{\begin{matrix}
-\sum_{(u,i,j)\in\mathcal{O} }^{} log(\sigma (\bigtriangleup S_k^* )), k=1\\
\mathcal{L}^{'}_{boost},k>1.
\end{matrix}\right.
\end{equation}

The output of the cumulative model is then updated by adding the predicted scores of the current model and the previous cumulative model. As a result, the updated cumulative model moves in the direction that most rapidly reduces the loss.

\subsubsection{Distillation Step}
Traditional GB constructs the final cumulative model by necessitating the storage of all model weights from previous epochs, which results in increased memory demands and conflicts with the prevailing deep learning paradigm of retaining only the latest model weights.
To this end, we design a distillation step to update the cumulative model in a space-efficient manner after the boosting step. Specifically, the previous cumulative model and the current model are treated as teacher models; the cumulative model being updated serves as the student model. The student model aims to approximate the sum of score differences predicted by two teacher models. The corresponding objective function can be expressed as:
\begin{equation}
\mathcal{L}_{distill}=\sum_{(u,i,j)\in\mathcal{O}}^{} \left ( \bigtriangleup S_k-\texttt{sg}(\bigtriangleup S_{k-1}+\bigtriangleup S_k^*)\right )^2,
\end{equation}
where $\bigtriangleup S_k$ is the score difference predicted by the updated cumulative model of epoch $k$, and $(u, i,j)$ is omitted for brevity. In this way, the updated cumulative model inherits the exclusive discriminative information from previous historical models without explicitly storing all historical model weights.

In addition, the sampled model is trained greedily to fit the residuals during the boosting step, which makes the cumulative model prone to poor local minima \cite{ReconBoost}. Therefore, we introduce a global correction regularization to promote the optimality of the updated cumulative model, denoted as:
\begin{equation}
    \mathcal{L}_{gc}=-\sum_{(u,i,j)\in\mathcal{O} }^{} log(\sigma (\bigtriangleup S_k )),
\end{equation}
where $(u, i,j)$ is omitted for brevity. The overall loss function of the distillation step is:
\begin{equation}
\label{eq:ld}
    \mathcal{L}_{d}=\mathcal{L}_{gc}+\mathcal{L}_{distill}.
\end{equation}

The updated cumulative model thus learns in the direction of both fitting residual discriminative information and improving the overall performance.

\begin{algorithm}[t]
\caption{Training process of D3ER}
\label{algo}
\begin{algorithmic}[1]
\STATE \textbf{Input:} Raw multi-modal features $\mathcal{X}$, pretrain epochs $N$ for FCD, ensemble epochs $K$, learning rate $lr$, hyper-parameters $\alpha_c,\alpha_w, \{d_m| m\in\mathcal{M}\}$.
\STATE \textbf{Output:} The final cumulative model $\mathcal{G}_{\varTheta_{K}}=\left \{ g_{\vartheta_K^{c}}, g_{\vartheta_{v,K}^{s}},g_{\vartheta_{t,K}^{s}}\right \}$

\STATE \textit{\# Pre-train Stage}
\STATE \textbf{Initialize:} The network parameter of feature extractors for two modalities $\mathcal{F}_\theta=\left \{ f_{\theta_v^c},f_{\theta_v^s},f_{\theta_t^c},f_{\theta_t^s}\right \} $
\FOR{$n=1$ to $N$}
\STATE \textit{\# Train the feature extractors to acquire HOI and HEI}
\STATE $\theta\leftarrow\theta-lr \cdot \triangledown_{\theta}\mathcal{L}_{FCD}(\mathcal{F}_\theta(\mathcal{X}) )$
\ENDFOR

\STATE \textit{\# Ensemble Stage}
\STATE \textbf{Initialize:} The pre-trained feature extractors are used to initialize the feature extraction part of the cumulative model, while the embedding encoders are initialized randomly, thereby initializing the cumulative model as $\mathcal{G}_{\varTheta_{0}}=\left \{g_{\vartheta_0^{c}}, g_{\vartheta_{v,0}^{s}},g_{\vartheta_{t,0}^{s}}\right \}$.
\FOR{$k=1$ to $K$}
\STATE \textit{\# Boosting step}
\STATE Sample a model $g_{\vartheta_{k-1}^*}$ from $\mathcal{G}_{\varTheta_{k-1}}$
\STATE \textit{\# Train the current model to learn the residual discriminative information}
\STATE $\vartheta_{k}^{*} \leftarrow \vartheta_{k-1}^{*}-lr \cdot \triangledown_{\vartheta_{k-1}^{*}}\mathcal{L}_{boost}(g_{\vartheta_{k-1}^{*}}(\mathcal{X}), \mathcal{G}_{\varTheta_{k-1}}(\mathcal{X}))$
\STATE \textit{\# Distillation step}
\STATE \textbf{Initialize:} initialize $\varTheta_{k}$ with $\varTheta_{k-1}$
\STATE \textit{\# Update the cumulative model by inheriting the knowledge of both the previous cumulative model and the current boosting model}

\STATE $\varTheta_{k} \leftarrow \varTheta_{k}-lr \cdot \triangledown_{\varTheta_{k}}\mathcal{L}_d(\mathcal{G}_{\varTheta_{k}}(\mathcal{X}),\mathcal{G}_{\varTheta_{k-1}}(\mathcal{X})+g_{\vartheta_{k}^{*}}(\mathcal{X}))$

\STATE Discard $\mathcal{G}_{\varTheta_{k-1}}$ and retain $\mathcal{G}_{\varTheta_{k}}$
\ENDFOR
\STATE \textbf{Return:} $\mathcal{G}_{\varTheta_{K}}$
\end{algorithmic}
\end{algorithm}
\subsection{Training Pipeline}
The detailed algorithm pseudo-code can be found in Algorithm \ref{algo}. Our D3ER optimizes disentanglement and ensemble in two sequential stages. Concretely, the FCD module is first pre-trained for $N$ epochs using Eq. (\ref{eq:fcd}). The cumulative model is then initialized with the pre-trained weights and optimized for $K$ epochs, guided by Eq. (\ref{eq:boost}) and Eq. (\ref{eq:ld}). The final cumulative model is then saved for top-$n$ recommendations.
\section{Theoretical Analysis}
\label{sec:thero}
In this section, we analyze the effectiveness of our proposed modules from a theoretical perspective.

According to the forward process of the proposed D3ER, we first confirm the validity of leveraging Wasserstein barycenter to obtain HOI from a target error \cite{courty2017joint,liu2021transfer,cao2022otkge} perspective.
Let $\mathcal{Z},\mathcal{Y}$ be the spaces where learned representations contain HOI and recommendation labels. Notice that $z_v^c,z_t^c$ and $z^c$ are assumed to obey the same true mapping function \( f^* : \mathcal{Z} \rightarrow \mathcal{Y} \) in MR tasks.
Considering the true mapping function \( f^* \) is unknown, we choose a function \( f \) from the hypothesis class \( \mathcal{F} \) for substitution, i.e., \( \forall f \in \mathcal{F}, f : \mathcal{Z} \rightarrow \mathcal{Y} \).

Obviously, there exists an approximation error between the hypothesis \( f \) and the true mapping function \( f^* \) under the distribution \( \mathcal{P} \), we measure the approximation error here with $\xi_{\mathcal{P}}(f, f^*) = \mathbb{E}_{z \in \mathcal{P}} [|f(z) - f^*(z)|] $, which is termed target error.
For the sake of conciseness, we denote $\xi_{\mathcal{P}}(f, f^*)$ by its abbreviation $\xi_{\mathcal{P}}(f)$.
Above all, we can derive the target error of the Wasserstein barycenter-based HOI from the following theorem:

\begin{theorem} \label{theorem:1}
\textbf{(Uniformity of Wasserstein barycenter-based HOI)}
We denote by $\mathcal{P}_{z_v^c}$, $\mathcal{P}_{z_t^c}$, and $\mathcal{P}_{z^c}$ for the distributions of $z_v^c,z_t^c$ and $z^c$. Assume the hypotheses $f, f^* \in \mathcal{F}$ are all $K$-Lipschitz continuous for some $K$. Then the following statement holds for every hypothesis $f, f^* \in \mathcal{F}$:
\begin{equation}
   \xi_{z^c}(f) \leq \mathop{min}_{m \in \{t,v\}} \{\xi_{z^c_m}(f) + \mathcal{W}(\mathcal{P}_{z_m^c}, \mathcal{P}_{z^c})\}.
\end{equation}
\end{theorem}

The $\mathcal{W}(\mathcal{P}_{z_m^c}, \mathcal{P}_{z^c})$ \textit{\textbf{cannot}} be minimized by archetypal MR approaches. Conversely, our method attains theoretically minimized $\mathcal{W}(\mathcal{P}_{z_m^c}, \mathcal{P}_{z^c})$ by $\mathcal{L}_{wd}$ and Eq. (\ref{eq:wdbf}). Furthermore, minimizing $\mathcal{W}(\mathcal{P}_{z_m^c},\allowbreak  \mathcal{P}_{z^c})$ and fusion by Eq. (\ref{eq:wdbf}) enable the construction of a coherent and unified $z^c$, wherein the post-fusion error is demonstrably lower than the error associated with any individual $z_m^c$. 
Then, we demonstrate the effectiveness of the KDBoost module from the generalization error perspective through Theorem \ref{theorem:2}.

\begin{theorem} \label{theorem:2}
(\textbf{Generalization Error of KDBoost}) Given the set of recommendation predictions $\hat{Y}_g$ ($g \in \mathcal{G}_{\Theta}$ and $\mathcal{G}_\Theta =\left \{ g_{\vartheta^c}, g_{\vartheta_v^s}, g_{\vartheta _t^s} \right \}$) and the loss function $\ell(\cdot)$. Then the following equation holds:
{\small
\begin{equation} \label{eq:theorem 2}
\begin{aligned}
\mathbb{E}_{\mathcal{D}}[\mathbb{E}_{\mathcal{O},\mathcal{Y}}[&\ell(\overline{Y},Y)]]= \mathbb{E}_{\mathcal{O}}[\underbrace{\mathbb{E}_{\mathcal{Y}|\mathcal{O}}[\ell(Y^*,Y)]}_{Noise}+\underbrace{\frac{1}{|\mathcal{G}_{\Theta}|}\sum_{g 
\in \mathcal{G}_{\Theta}}\ell(\tilde{Y}_g,Y^*)}_{\textcolor{red}{Average \; bias}}\\
+&\underbrace{\frac{1}{|\mathcal{G}_{\Theta}|}\sum_{g 
\in \mathcal{G}_{\Theta}}\mathbb{E}_{\mathcal{D}}[\ell(\hat{Y}_g,\tilde{Y}_g)]}_{Average \; variance} 
-\underbrace{\mathbb{E}_{\mathcal{D}}[\frac{1}{|\mathcal{G}_{\Theta}|}\sum_{g 
\in \mathcal{G}_{\Theta}}\ell(\hat{Y}_g,\overline{Y})]]}_{\textcolor{red}{Diversity}},
\end{aligned}
\end{equation}}
\end{theorem}
where $\mathbb{E}(\cdot)$ is the empirical error, $Y^*=\mathbb{E}_{\mathcal{Y}|\mathcal{O}}[Y]$ is the Bayes-optimal prediction, $\overline{Y}=\underset{Y\in\mathcal{Y}}{\operatorname*{\operatorname*{\arg\min}}}\frac{1}{|\mathcal{G}_{\Theta}|}\sum_{g 
\in \mathcal{G}_{\Theta}}\ell(\hat{Y}_g,Y)$ is the minimizer of the averaged loss $\ell(\hat{Y}_g, Y)$ over all recommendation models in $\mathcal{G}_{\Theta}$, and $\tilde{Y}_i\overset{def}{\operatorname*{=}}\arg\min_{Y\in\mathcal{Y}}\mathbb{E}_\mathcal{D}[\ell(\hat{Y}_i,Y)]$. The proposed KDBoost module reduces the generalization error by the \textcolor{red}{red} term in Eq. (\ref{eq:theorem 2}), i.e., diminishing the average bias and enhancing the diversity term. Specifically, it adopts an alternating learning strategy to boost the performance of each model $g \in \mathcal{G}_{\Theta}$, thereby lowering the average bias term. Moreover, KDBoost optimizes model $g$ to learn the discriminative information that remains unacquired by $g' \in \mathcal{G}_{\Theta}/g$. This optimization objective leads different models to focus on distinct discriminative information, which promotes the diversity of models in $\mathcal{G}_{\Theta}$. Consequently, D3ER can achieve a tighter generalization error bound compared to conventional MR models.

\begin{table*}
\centering
\tabcolsep=0.11cm
\caption{Quantitative comparisons with several baselines on three datasets. The best results and second-best results are marked in \textcolor{red}{red} and \textcolor{blue}{blue}, respectively. ``*'' means the improvement is statistically significant with $p$-value $<$ 0.05 in paired t-tests.}
\label{tab:comparison}
\begin{tabular}{lcccccccccccc}%
\toprule 
\multirow{2}{*}{\textbf{Method}} & \multicolumn{4}{c}{\textbf{Baby}} & \multicolumn{4}{c}{\textbf{Sports}} & \multicolumn{4}{c}{\textbf{Clothing}} \\  \cmidrule(lr){2-5} \cmidrule(lr){6-9} \cmidrule(lr){10-13}
& \textbf{R@20} & \textbf{R@50} & \textbf{N@20} & \textbf{N@50} & \textbf{R@20} & \textbf{R@50} & \textbf{N@20} & \textbf{N@50} & \textbf{R@20} & \textbf{R@50} & \textbf{N@20} & \textbf{N@50} \\
\midrule
MF-BPR \cite{MF-BPR} & 0.0562 & 0.0995 & 0.0240 & 0.0327 & 0.0648 & 0.1071 & 0.0282 & 0.0367 & 0.0276 & 0.0443 & 0.0128 & 0.0161 \\
LightGCN \cite{LightGCN} & 0.0744 & 0.1308 & 0.0320 & 0.0432& 0.0841 & 0.1396 & 0.0380 & 0.0490 & 0.0514 & 0.0817 & 0.0231 & 0.0282\\
LayerGCN \cite{zhou2023layer} & 0.0806 & 0.1419 & 0.0342 & 0.0464 & 0.0899 & 0.1468 & 0.0396 & 0.0510 & 0.0551 & 0.0886 & 0.0247 & 0.0313 \\
\midrule
VBPR \cite{VBPR} & 0.0308 & 0.0537 & 0.0128 & 0.0174 & 0.0310 & 0.0522 & 0.0131 & 0.0179 & 0.0228 & 0.0041 & 0.0096 & 0.0132 \\
MMGCN \cite{MMGCN} & 0.0589 & 0.1110 & 0.0240 & 0.0344 & 0.0559 & 0.0978 & 0.0238 & 0.0322 & 0.0336 & 0.0604 & 0.0138 & 0.0191 \\
LATTICE \cite{LATTICE} & 0.0834 & 0.1455 & 0.0360 & 0.0485 & 0.0920 & 0.1496 & 0.0407 & 0.0522 & 0.0733 & 0.1142 & 0.0329 & 0.0411 \\
SLMRec \cite{SLMRec} & 0.0746 & 0.1222 & 0.0331 & 0.0426 & 0.0947 & 0.1513 & 0.0438 & 0.0551 & 0.0657 & 0.1069 & 0.0295 & 0.0377 \\
FREEDOM \cite{zhou2023tale} & 0.0866 & 0.1545 & 0.0371 & 0.0506 & 0.0944 & 0.1555 & 0.0419 & 0.0514 & 0.0863 & 0.1378 & 0.0382 & 0.0484 \\
BM3 \cite{BM3} & 0.0833 & 0.1437 & 0.0342 & 0.0463 & 0.0860 & 0.1446 & 0.0380 & 0.0497 & 0.0569 & 0.0908 & 0.0252 & 0.0320 \\
MMSSL \cite{MMSSL} & 0.0932 & 0.1572 & 0.0415 & \textcolor{blue}{0.0548} & 0.0992 & 0.1581 & 0.0460 & 0.0582 & 0.0750 & 0.1197 & 0.0344 & 0.0436 \\
AlignRec \cite{liu2024alignrec} & 0.0759 & 0.1320 & 0.0330 & 0.0442 & 0.0917 & 0.1473 & 0.0408 & 0.0519 & 0.0652 & 0.1064 & 0.0286 & 0.0369 \\
LGMRec \cite{guo2024lgmrec} & 0.0943 & 0.1599 & 0.0406 & 0.0538 & 0.0982 & 0.1575 & 0.0440 & 0.0559 & 0.0822 & 0.1306 & 0.0367 & 0.0463   \\
\midrule
MGCN \cite{yu2023multi} & 0.0904 & 0.1491 & 0.0392 & 0.0510 & 0.1013 & 0.1631 & 0.0454 & 0.0578 & 0.0959 & 0.1462 & 0.0435 & 0.0535 \\
\rowcolor{gray!10}
\textbf{MGCN+D3ER} & \textbf{0.0953*} & \textbf{0.1604*} & \textbf{0.0411*} & \textbf{0.0541*} & \textbf{0.1071*} & \textcolor{blue}{\textbf{0.1700*}} & \textbf{0.0480*} & \textbf{0.0606*} & \textbf{0.1001*} & \textbf{0.1529*} & \textbf{0.0453*} & \textbf{0.0559*}  \\

\textit{Improv.} & \textit{5.42\%} & \textit{7.58\%} & \textit{4.85\%} & \textit{6.08\%} & \textit{5.73\%} & \textit{4.23\%} & \textit{5.73\%} & \textit{4.84\%} & \textit{4.38\%} & \textit{4.58\%} & \textit{4.14\%} & \textit{4.49\%} \\
\midrule
DiffMM \cite{DiffMM} & \textcolor{blue}{0.0975} & \textcolor{blue}{0.1633} & 0.0411 & 0.0540 & 0.1017 & 0.1656 & 0.0458 & 0.0581 & 0.0798 & 0.1259 & 0.0354 & 0.0446  \\
\rowcolor{gray!10}
\textbf{DiffMM+D3ER} & \textcolor{red}{\textbf{0.1018*}} & \textcolor{red}{\textbf{0.1692*}} & \textcolor{red}{\textbf{0.0431*}} & \textcolor{red}{\textbf{0.0564*}} & \textbf{0.1062*} & \textbf{0.1680*} & \textbf{0.0472*} & \textbf{0.0596*} & \textbf{0.0849*} & \textbf{0.1344*} & \textbf{0.0374*} & \textbf{0.0474*} \\

\textit{Improv.} & \textit{4.41\%} & \textit{3.61\%} & \textit{4.87\%} & \textit{4.44\%} & \textit{4.42\%} & \textit{1.45\%} & \textit{3.06\%} & \textit{2.58\%} & \textit{6.39\%} & \textit{6.75\%} & \textit{5.65\%} & \textit{6.28\%}  \\
\midrule
PGL \cite{PGL} & 0.0930 &  0.1552 & 0.0404 & 0.0529 & \textcolor{blue}{0.1078} & 0.1687 & \textcolor{blue}{0.0486} & \textcolor{blue}{0.0608} & \textcolor{blue}{0.1037} & \textcolor{blue}{0.1599} & \textcolor{blue}{0.0467} & \textcolor{blue}{0.0579} \\
\rowcolor{gray!10}
\textbf{PGL+D3ER} & \textbf{0.0961*} & \textbf{0.1607*} & \textcolor{blue}{\textbf{0.0417*}} & \textbf{0.0547*} & \textcolor{red}{\textbf{0.1124*}} & \textcolor{red}{\textbf{0.1749*}} & \textcolor{red}{\textbf{0.0509*}} & \textcolor{red}{\textbf{0.0634*}} & \textcolor{red}{\textbf{0.1058*}} & \textcolor{red}{\textbf{0.1654*}} & \textcolor{red}{\textbf{0.0478*}} & \textcolor{red}{\textbf{0.0595*}} \\

\textit{Improv.} & \textit{3.33\%} & \textit{3.54\%} & \textit{3.22\%} & \textit{3.40\%} & \textit{4.27\%} & \textit{3.68\%} & \textit{4.73\%} & \textit{4.28\%} & \textit{2.03\%} & \textit{2.88\%} & \textit{2.36\%} & \textit{2.76\%}  \\
\bottomrule
 
\end{tabular}
\end{table*}

\begin{table}
\tabcolsep=0.1cm
\small
\centering
\caption{Ablation studies of various components.}
\label{tab:ablation}
\begin{tabular}{l|cc|cc|cc}
\toprule
\multirow{2}{*}{\textbf{Variants}} & \multicolumn{2}{c|}{\textbf{Baby}} & \multicolumn{2}{c|}{\textbf{Sports}} & \multicolumn{2}{c}{\textbf{Clothing}} \\

& \textbf{R@20} & \textbf{N@20} & \textbf{R@20} & \textbf{N@20} & \textbf{R@20} & \textbf{N@20}\\
\midrule
D3ER$_{w/o~cl}$   & 0.0986 & 0.0421 & 0.1045 & 0.0467 & 0.0814 & 0.0359 \\
D3ER$_{w/o~wd}$   & 0.0997 & 0.0421 & 0.1044 & 0.0467 & 0.0831 & 0.0361 \\
D3ER$_{w/o~fcd}$   & 0.0978 & 0.0424 & 0.1023 & 0.0456 & 0.0809 & 0.0361 \\
D3ER$_{w/o~b\&d}$ & 0.0957 & 0.0408 & 0.0926 & 0.0409 & 0.0654 & 0.0285 \\
D3ER$_{w/o~gc}$   & 0.0664 & 0.0266 & 0.0715 & 0.0302 & 0.0539 & 0.0224 \\
\midrule
\rowcolor{gray!10}
\textbf{D3ER} & \textbf{0.1018} & \textbf{0.0431} & \textbf{0.1062} & \textbf{0.0472} & \textbf{0.0849} & \textbf{0.0374} \\
\bottomrule
\end{tabular}

\vspace{-0.5cm}
\end{table}

\section{Experiments}
\label{sec:expri}

\subsection{Experimental Setup}
\subsubsection{Evaluation Datasets} We evaluate the performance of our proposed D3ER on three widely used real-world datasets, sourced from the Amazon review dataset \cite{Amazon}. These datasets consist of products in the categories: baby, sports and outdoors, clothing and shoes, respectively. For brevity, we refer to them as the Baby, Sports, and Clothing dataset.
Following \cite{DiffMM}, the raw visual features of the Baby and Sports datasets are obtained by pretrained CLIP-ViT \cite{radford2021learning} model. For the clothing dataset, consistent with \cite{BM3}, visual features are extracted from VGG-16 \cite{simonyan2014very} model. For raw textual features, they are extracted through the pretrained Sentence-BERT \cite{reimers2019sentence} for all three datasets.

\subsubsection{Metrics}Different model performances of top-$n$ recommendation are evaluated using two metrics: Recall@$n$ (R@$n$) and Normalized Discounted Cumulative Gain (N@$n$) \cite{he2015trirank}. In our paper, we report the results with $n$ equal to 20 and 50.

\subsubsection{Comparison Baselines} We compare our D3ER with the following typical models: 1) General recommendation models: MF-BPR \cite{MF-BPR}, LightGCN \cite{LightGCN}, LayerGCN \cite{zhou2023layer}; 2) Multi-modal recommendation models: VBPR \cite{VBPR}, MMGCN  \cite{MMGCN}, LATTICE \cite{LATTICE}, SLMRec \cite{SLMRec}, FREEDOM \cite{zhou2023tale}, BM3 \cite{BM3}, MMSSL \cite{MMSSL}, AlignRec \cite{liu2024alignrec}, LGMRec \cite{guo2024lgmrec}, MGCN \cite{yu2023multi}, DiffMM \cite{DiffMM}, and PGL \cite{PGL}.

\subsubsection{Implementation Details} 

We develop our D3ER using the PyTorch framework and conduct experiments on an RTX 3090 GPU. Note that the embedding encoder can be any CF model, here, we employ three popular multi-modal recommendation models as embedding encoders, namely MGCN \cite{yu2023multi}, DiffMM \cite{DiffMM}, and PGL \cite{PGL}. All models are trained using the Adam optimizer, with the learning rate set to 1e-3. For all datasets, $p$ for obtaining $\mathcal{L}_{wd}$, disentanglement epoch $N$, and ensemble epoch $K$ are set to 2, 10, and 50, respectively. Besides, the value of hyper-parameters is adjusted to fit the characteristics of each dataset. Specifically, $\alpha_c$ and $\alpha_w$ are searched in \{1e-4,1e-3,1e-2,1e-1\}, and $d_m$ is searched in \{0.01,0.1,1,10\}.

\subsection{Performance Comparisons}
The performance evaluation of our proposed D3ER is demonstrated in Table \ref{tab:comparison}, with the following key observations: Firstly, compared to 15 recommendation models, our method consistently achieves state-of-the-art recommendation performance across all datasets. Specifically, with the powerful backbone PGL as the embedding encoder, our D3ER outperforms all compared methods on the Sports and Clothing datasets. When integrated into the advanced DiffMM model, our method yields the highest Recall and NDCG on the Baby dataset. 

Secondly, our D3ER framework delivers consistent and statistically significant performance gains across three distinct recommendation architectures. Notably, it achieves a 7.58\% improvement in Recall@50 over the backbone MGCN. Recent approaches (e.g., BM3, MMSSL, and DiffMM) employ contrastive learning to align multi-modal representations and acquire HOI, overlooking valuable HEI. MGCN focuses on modality fusion but neglects the importance of optimizing strategies for diverse information. In contrast, D3ER advances recommendation accuracy by sufficiently disentangling and effectively ensembling modal-homogeneity and heterogeneity features, validating the advantage of our design in MR systems.

\subsection{Ablation Study}
To investigate the effect of several key designs, we conduct ablation studies with DiffMM \cite{DiffMM} employed as backbone, which are presented in Table \ref{tab:ablation}. 

\subsubsection{Impact of FCD} To validate the necessity of the disentangling process, we remove the FCD module, with the resultant model denoted as ``D3ER$_{w/o\,fcd}$''. 
Since the loss terms associated with FCD are only active during the pre-training phase, this variant is implemented by setting $N$ to 0.
The consistent performance drop indicates that explicitly decoupling HOI and HEI through our FCD module facilitates the learning of informative item representations, thereby enhancing recommendation performance.

\subsubsection{Impact of Loss Terms in Inter-Modal Alignment} To investigate the roles of the instance-level $\mathcal{L}_{cl}$ and the distribution-level $\mathcal{L}_{wd}$ in inter-modal alignment, we design two variants, each omitting one of the two losses, denoted as ``D3ER$_{w/o\,cl}$'' and ``D3ER$_{w/o\,wd}$'' in Table \ref{tab:ablation}, respectively. The results indicate that removing either alignment loss leads to performance degradation, emphasizing their unique contribution. Employing both losses simultaneously yields the best performance, illustrating that instance-level and distribution-level alignment can work together for more sufficient disentanglement.

\subsubsection{Impact of KDBoost} To assess the contribution of our KDBoost in ensembling models containing different types of information, we remove $\mathcal{L}_{boost}$ and $\mathcal{L}_{distill}$, retaining only the global correction regularization term for alternating training. This variant is denoted as ``D3ER$_{w/o\,b\&d}$''. Significant performance drops are observed across all three datasets, highlighting the critical role of 
$\mathcal{L}_{boost}$ and $\mathcal{L}_{distill}$ in learning and transferring residual discriminative knowledge, respectively.

\begin{figure*}[htbp]
    \centering
    
    \subfigure{\includegraphics[width=0.23\linewidth]{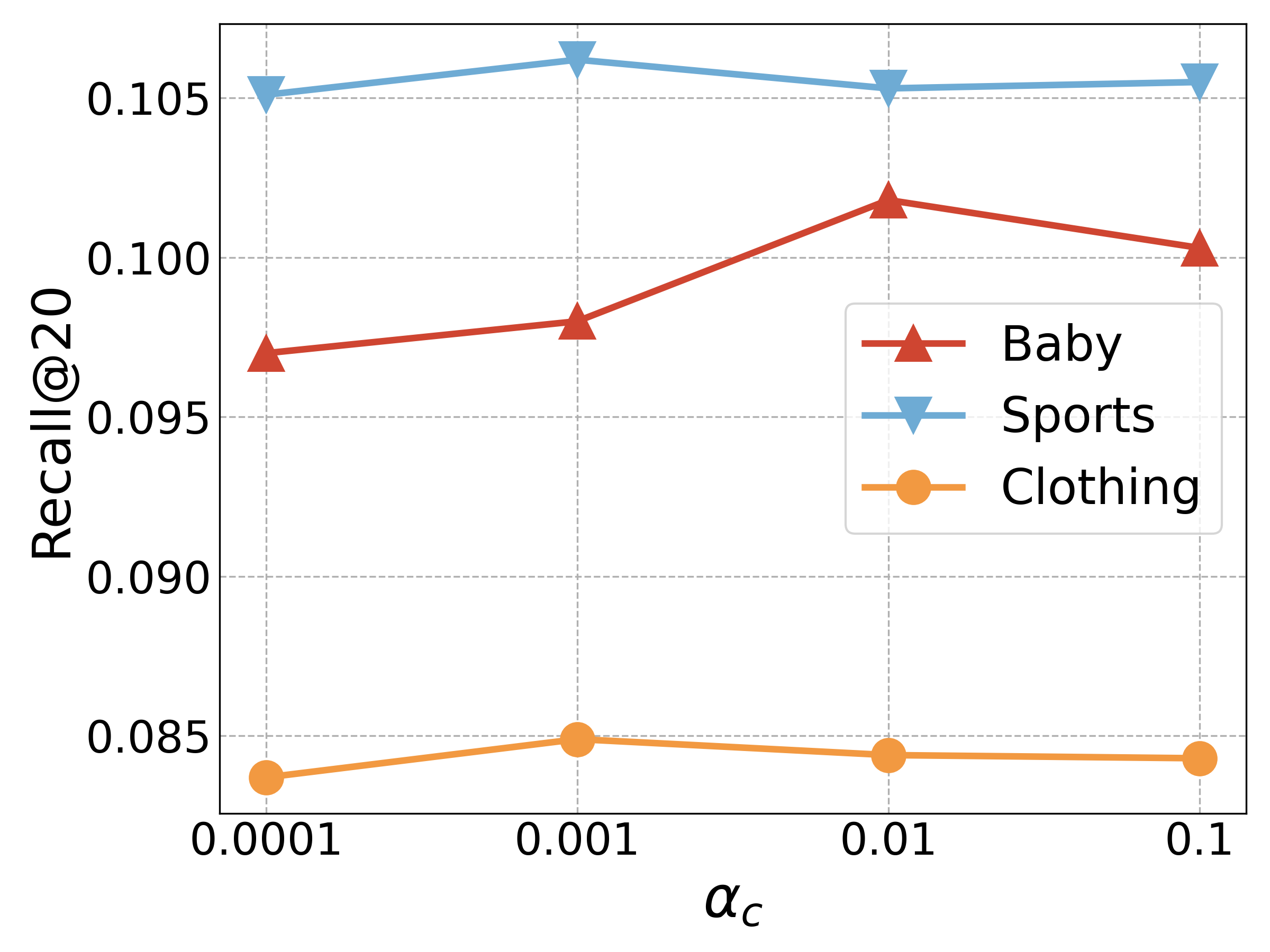}}
    \hfill
    \subfigure{\includegraphics[width=0.23\linewidth]{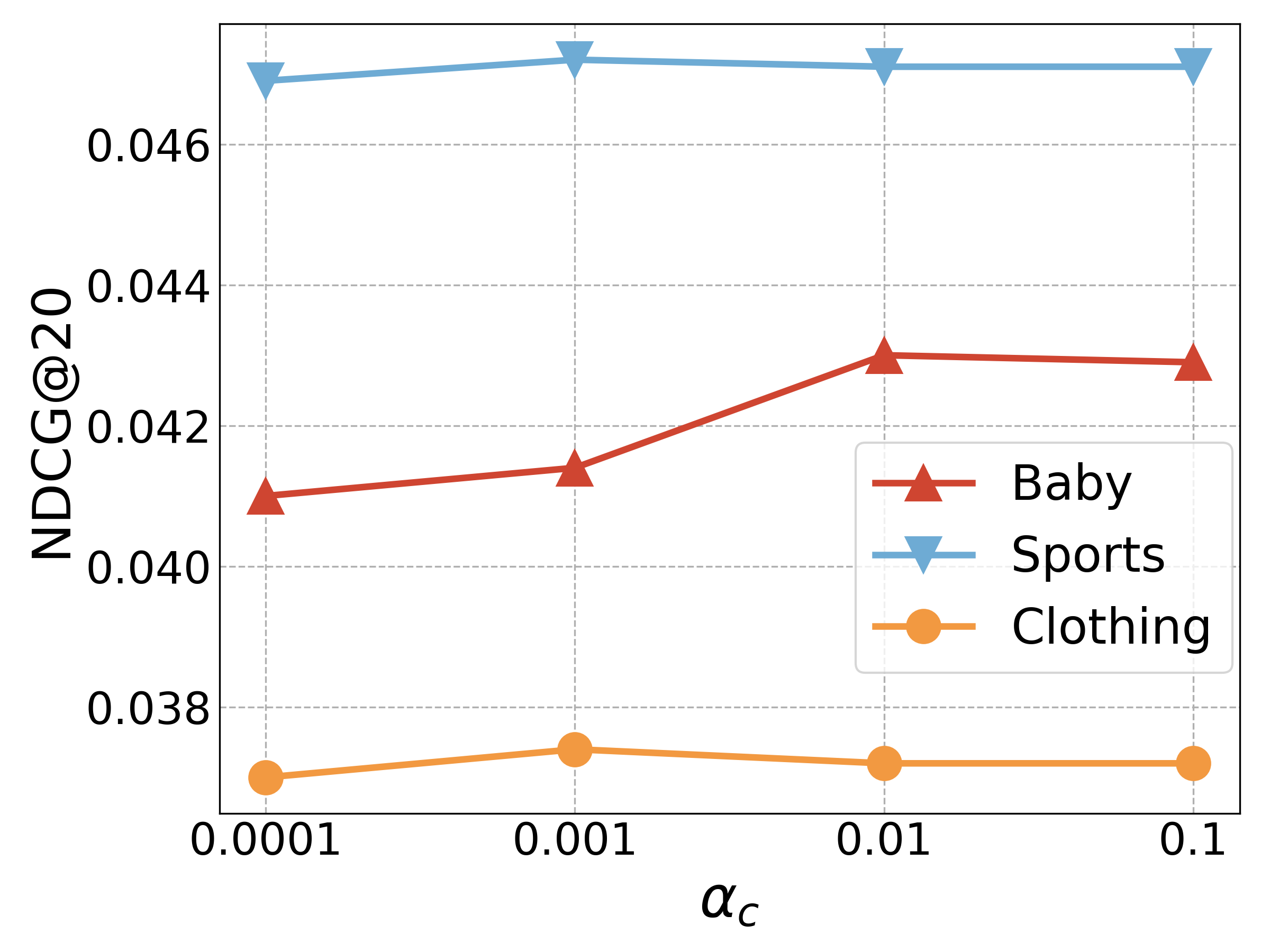}}
    \hfill
    \subfigure{\includegraphics[width=0.23\linewidth]{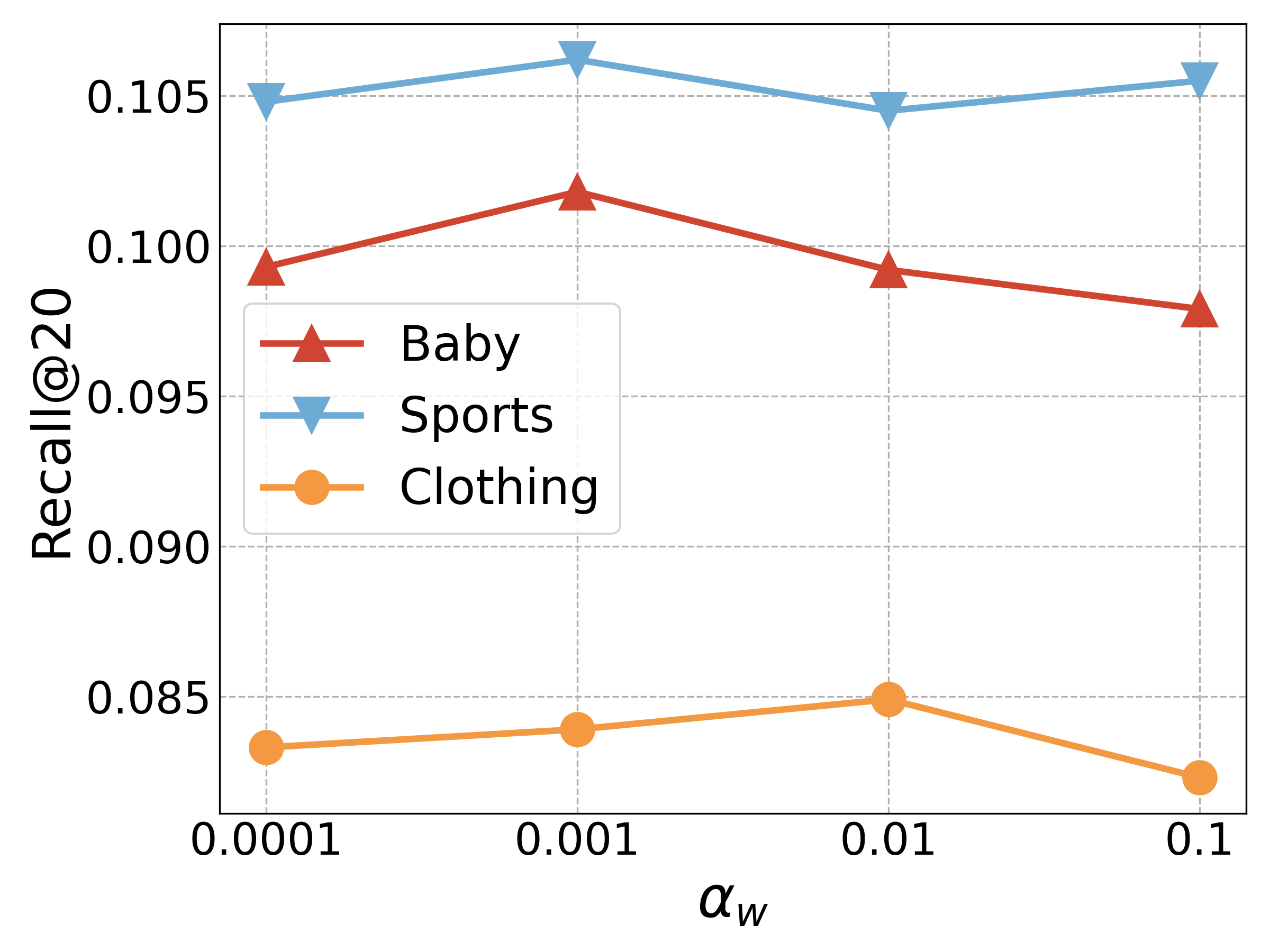}}
    \hfill
    \subfigure{\includegraphics[width=0.23\linewidth]{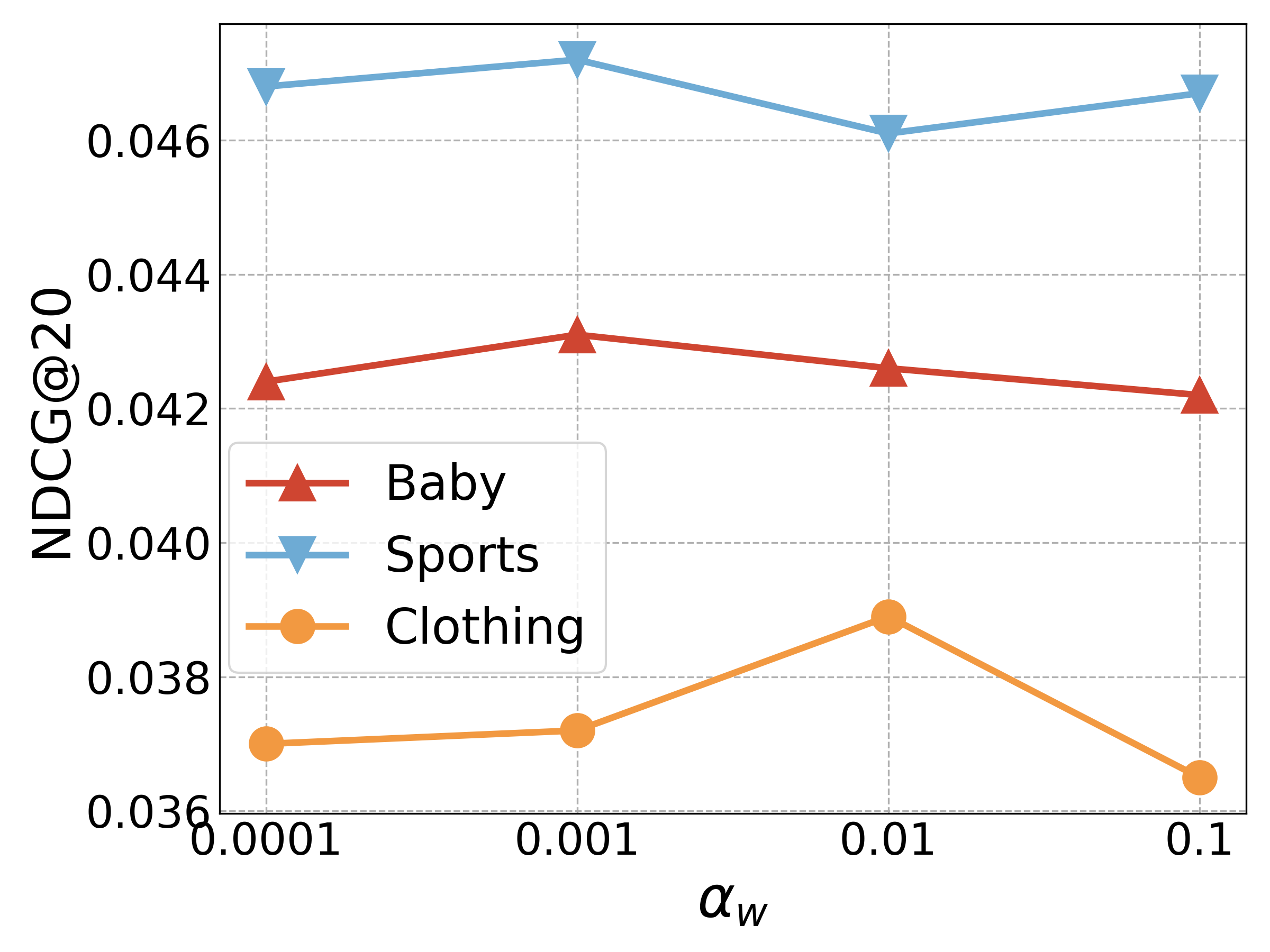}}
    \vspace{-0.1cm}
    \caption{Influence of loss weights $\alpha_c$ and $\alpha_w$ on Recall@20 and NDCG@20 over three datasets.}
    \label{fig_hyp1}
\end{figure*}

\begin{figure}[htbp]
    \centering
    \centering
    
    \subfigure[Baby dataset]{\includegraphics[width=0.45\linewidth]{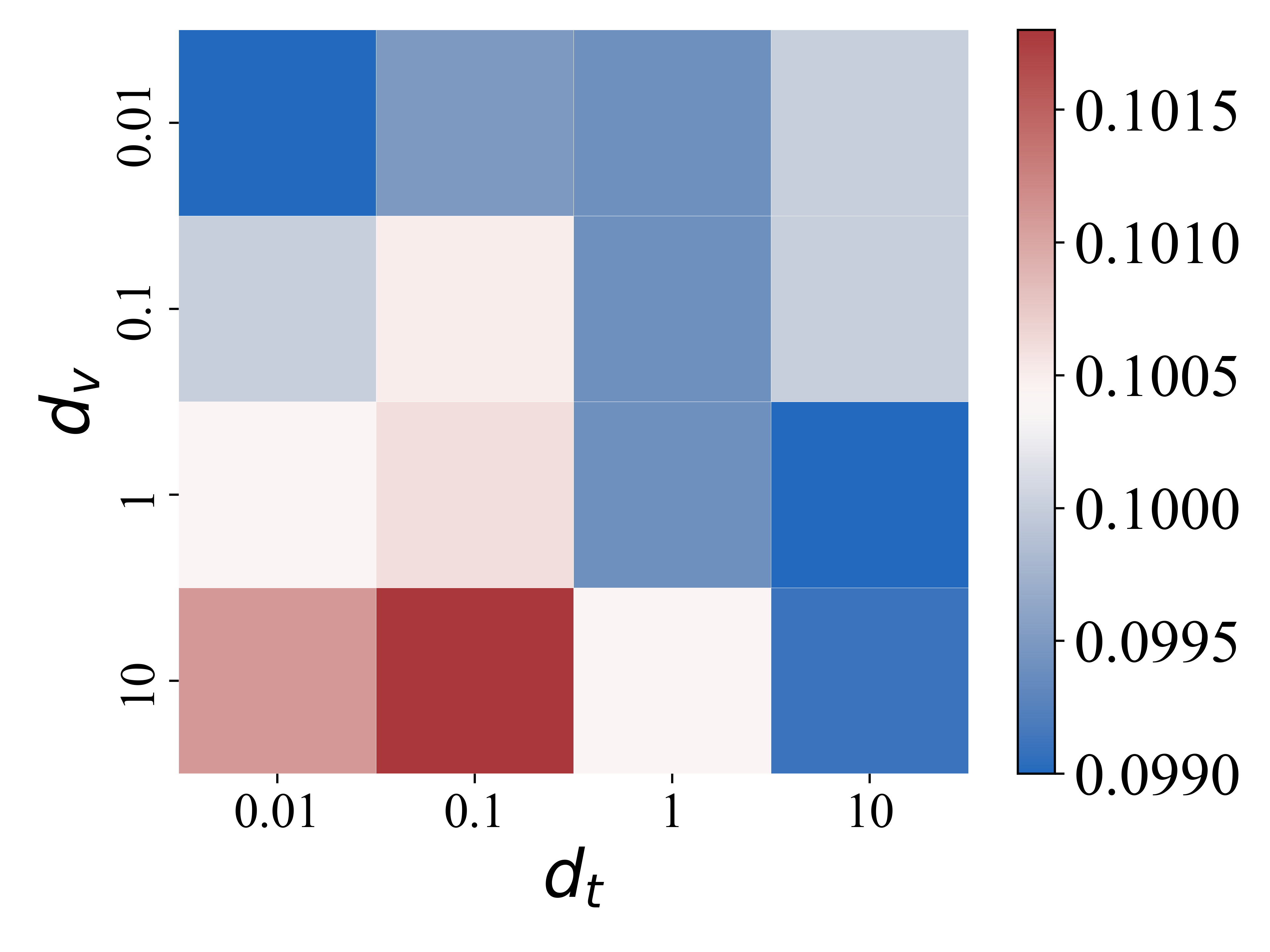}}
    \subfigure[Sports dataset]{\includegraphics[width=0.45\linewidth]{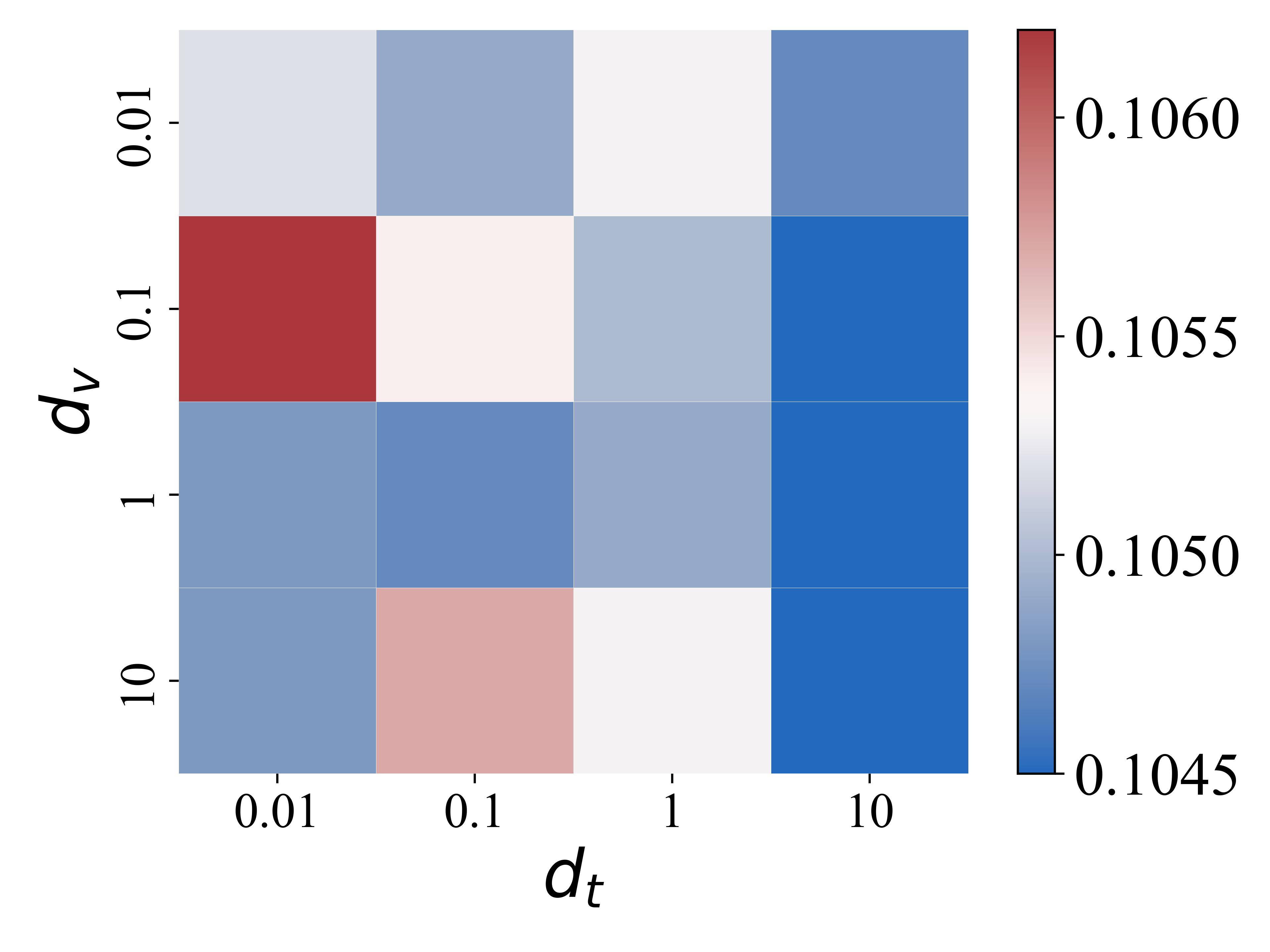}}
    
    \subfigure[Clothing dataset]{\includegraphics[width=0.45\linewidth]{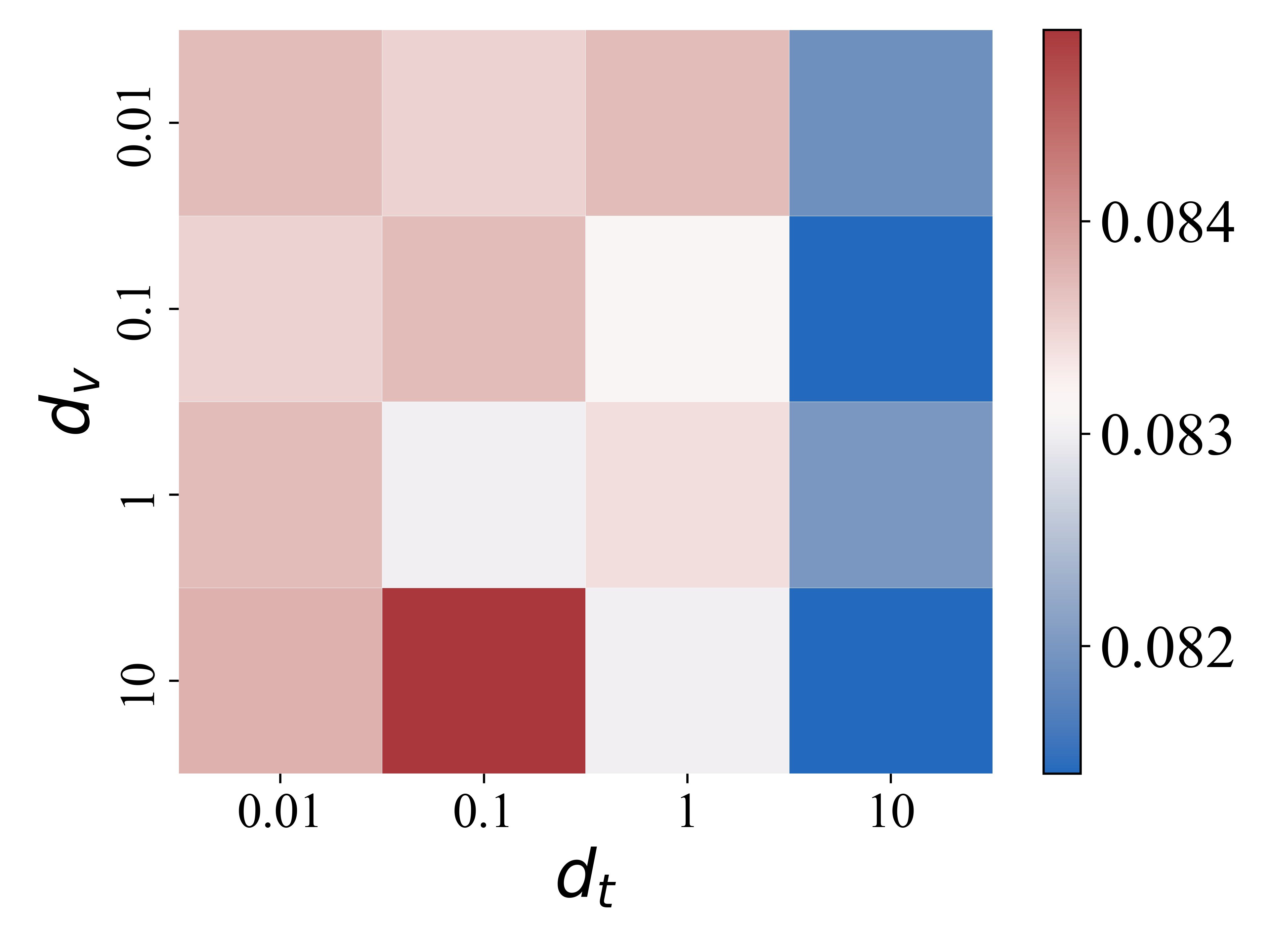}}
    \subfigure[t-SNE analysis]{\includegraphics[width=0.43\linewidth]{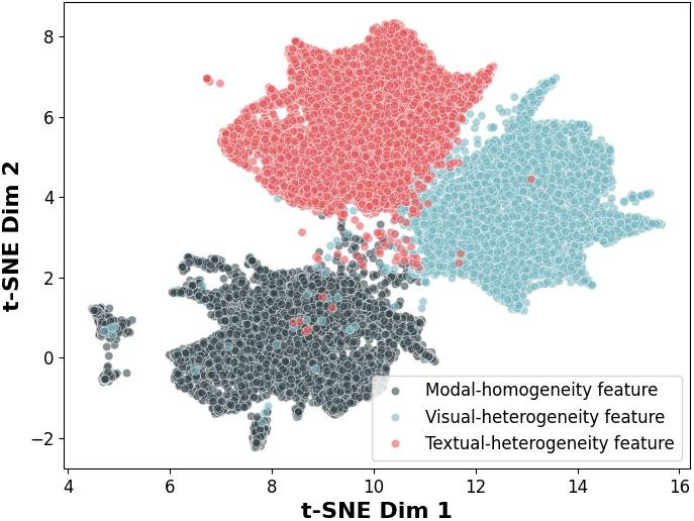}}
    \caption{Visualization and sensitivity studies on D3ER. Results are obtained by employing DiffMM as the backbone. (a-c) Recall@20 for varying combinations of threshold hyper-parameters $\{d_v,d_t\}$. (d) t-SNE visualization results of three types of disentangled features on Baby dataset.}
    \label{fig_study}
\end{figure}

\subsubsection{Impact of Global Correction Regularization} We further ablate the global correction regularization by removing $\mathcal{L}_{gc}$, resulting in the variant ``D3ER$_{w/o\,gc}$''. Its poorest performance among all variants validates the issue of local optima caused by the greedy learning nature of gradient boosting, underscoring the necessity of global regularization in ensemble training.

\subsection{Hyper-parameter and Visualization Analysis}

\subsubsection{Sensitivity of Weighting Hyperparameter} We present the impact of varying weight hyper-parameter values on the experimental results. As shown in Figure \ref{fig_hyp1}, the performance across all three datasets exhibits an initial increase followed by a decline as the weights $\alpha_c$ and $\alpha_w$ grow. Furthermore, the contributions of the two loss terms vary across datasets. These results underscore the necessity of weight selection in achieving instance-level and distribution-level alignment, thereby enhancing recommendation performance. This can be effectively accomplished by appropriately adjusting the weights within our provided hyper-parameter range.

\subsubsection{Sensitivity of Threshold Hyperparameter} Furthermore, we present sensitivity analyses of threshold hyper-parameters in D3ER. Figure \ref{fig_study}(a-c) shows the effect of different distance threshold combinations $\left \{ d_v,d_t \right \} $. It can be observed that too small thresholds fail to capture sufficient discriminative information, while overly large ones introduce noise into the features. Optimal threshold values vary across datasets and can be tuned via grid search to balance relevance and informativeness for a specific dataset.

\subsubsection{Visualization Analysis} To experimentally demonstrate the effect of separating HOI and HEI, we perform t-SNE visualization \cite{van2008visualizing} on the disentangled representations (i.e., $z^c$, $z_v^s$, and $z_t^s$). Specifically, after the model is fully trained, we project the disentangled features $z^c$, $z_v^s$, and $z_t^s$ into a two-dimensional space using t-SNE. The visualization results shown in Figure \ref{fig_study}(d), exhibit a clear separation among three types of features, supporting our motivation for disentangling HOI and HEI.

\begin{figure}
\centering
\includegraphics[width=\linewidth]{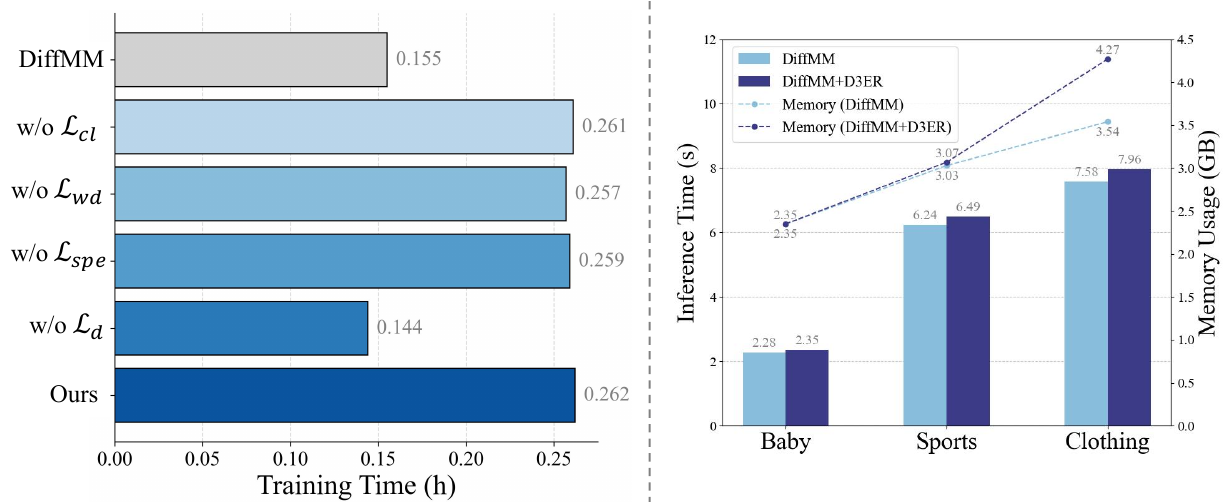}
\caption{Computational cost analysis.}
\label{fig_cost}
\end{figure}

\subsection{Efficiency Analysis}
To facilitate broader applicability of D3ER, we analyze the computational complexity as depicted in Figure \ref{fig_cost}. Overall, D3ER exhibits comparable inference speed and memory usage relative to the backbone model. To assess the cost of individual components, we ablate each loss term. The results reveal that the loss for disentanglement, operating only on item representations, introduces negligible training overhead. The primary additional training cost stems from the distillation step ($\mathcal{L}_{d}$), which is designed to reduce storage overhead and eliminate the need to store all historical model weights in traditional gradient boosting. For specific recommendation applications, researchers may adjust or simplify this step depending on their efficiency requirements.

\section{Conclusion}
We observe that HOI and HEI exhibit prediction preference to MR, and that the naive ensemble strategy cannot capture comprehensive sample-oriented discriminative information. To this end, we propose D3ER, which consists of FCD and KDBoost modules. The FCD module decouples item representations into modal-shared and modal-specific components through inter-modal alignment from the instance level and distribution level, as well as intra-modal separation. The recommendation models for each type of information are then alternately optimized by the KDBoost module through a boosting step and a distillation step, facilitating the learning of comprehensive sample-oriented discriminative information. The consistent performance improvement on three popular datasets has demonstrated the effectiveness of our method for MR.

\section*{Acknowledgments}
This work is supported by National Natural Science Foundation of China No. 62406313, Postdoctoral Fellowship Program of China Postdoctoral Science Foundation, Grant No. YJB20250283.

\bibliographystyle{ACM-Reference-Format}
\bibliography{ref}


\end{document}